\documentclass{acm_proc_article-sp}
\usepackage{syntax}
\usepackage{url}
\usepackage{graphics}
\usepackage[textsize=footnotesize,bordercolor=black!20]{todonotes}
\usepackage{microtype}
\usepackage[T1]{fontenc}
\usepackage[scaled]{beramono}
\usepackage{listings}

\definecolor{javared}{rgb}{0.6,0,0} 
\definecolor{javagreen}{rgb}{0.25,0.5,0.35} 
\definecolor{javapurple}{rgb}{0.5,0,0.35} 
\definecolor{javadocblue}{rgb}{0.25,0.35,0.75} 
 
\newcommand\Small{\fontsize{8}{8.0}\selectfont}
\newcommand*\LSTfont{\Small\ttfamily\SetTracking{encoding=*}{-60}\lsstyle}
\lstnewenvironment{java}[1][]
{\lstset{#1,
         frame=single, 
         captionpos=b,
         basicstyle=\LSTfont,
         escapeinside={$}{$}}}
{}

\newcommand{\metavar}{$\square$}

\begin{document}

\title{Identifying change patterns in software history}

%
%
%
%
%

\numberofauthors{2} 
%
\author{
%
%
\alignauthor
Jason Dagit\\
       \affaddr{Galois, Inc}\\
       \affaddr{Portland, OR}\\
       \email{dagitj@gmail.com}
\alignauthor
Matthew Sottile\\
       \affaddr{Galois, Inc}\\
       \affaddr{Portland, OR}\\
       \email{mjsottile@gmail.com}
}
\date{4 July 2013}

\maketitle
\begin{abstract}


Traditional algorithms for detecting differences in source code focus on
differences between lines.  As such, little can be learned about abstract
changes that occur over time within a project.  Structural differencing on the
program's abstract syntax tree reveals changes at the syntactic level within
code, which allows us to further process the differences to understand their
meaning. We propose that grouping of changes by some metric of similarity,
followed by pattern extraction via antiunification will allow us to identify
patterns of change  within a software project from the sequence of changes
contained within a Version Control System (VCS). Tree similarity metrics such
as a tree edit distance can be used to group changes in order to identify
groupings that may represent a single class of change (e.g., adding a parameter
to a function call). By applying antiunification within each group we are
able to generalize from families of concrete changes to patterns of structural
change.  Studying patterns of change at the structural level, instead of
line-by-line, allows us to gain insight into the evolution of software.

\end{abstract}




\section{Introduction}

Version control systems (VCS's) track the evolution of software over time in
the form of a sequence of changes to the plain text representation of the
code. We would like to be able to characterize the changes to files in a
software project according to the type of change that they represent.  The
ability to map these changes to the syntax of the language, instead of its raw
text representation, will allow them to be understood in terms of the language
constructs themselves.  Doing so will allow us to identify patterns of changes
at the abstract syntax level, separate from syntax neutral changes to the text
such as layout variations.  As a result, the interpretation of changes is
made unambiguous given the definition of the abstract syntax of the language.

Finding common patterns for the changes to a source file gives us the ability
to understand, at a higher level, what sorts of revisions are happening.
Detecting simple changes, such as semaphore handling changes in system-level
software, we may think to use a textual search tool, such as {\tt grep}, to
search the source code for functions related to semaphores. Such tools are
unable to easily identify more complex patterns though that have no single
textual representation, such as instances of semaphore handling calls being made
within conditionals where the format of the conditional can vary.  Structure
aware searching would be necessary in this case, as treating the program as
raw text ignores important syntactic structure.

In an even more complicated situation, a programmer may be faced with a
code base that they are unfamiliar with.  In this case, the programmer may not
know a-priori what kinds of structures are important to look for related to
a certain kind of change.  Here, we would like to use the differences that are
recorded in the VCS during the period of time when the change of interest was
being performed to discover the structural patterns that represent the high
level structure of the changes.  In this way, our goal is to not provide simply
a sophisticated search tool, but to provide a method for identifying patterns 
of code changes over a period of time.

Our contributions towards this goal presented in this paper are:

\begin{itemize}

\item We show that structural differencing algorithms that operate on the
abstract syntax tree (AST) of a language can be used to map text
differences stored in a VCS to a form where the syntactic meaning of changes
can be reasoned about.

\item We show that the antiunification algorithm that seeks the ``least general
generalization'' of a set of trees can be used to map changes considered to be
sufficiently similar to a meaningful generalized change pattern.

\item We show that a thresholded tree similarity metric derived from
a tree edit distance score provides a useful grouping mechanism to define
the notion of ``sufficiently similar''.

\end{itemize}

In this paper, we briefly describe the building blocks of our work and show
preliminary results of this methodology as applied to version control
repositories for open source projects available online.  The projects studied
in this paper are ANTLR\footnote{\url{https://github.com/antlr/antlr4}} and
Clojure\footnote{\url{https://github.com/clojure/clojure}}, both written in
Java.

\subsection{Motivation}
\label{sec:motivation}

We would like to be able to take existing software projects and use the history
stored in the VCS to answer questions which may be important to software
developers, software project managers, language designers, and static analysis
tools.

Consider a problem that has been faced by many projects in the last decade ---
the challenge of migrating to utilize multicore processors.  A manager who is
leading a large software project may want to answer important questions to
help inform future development: what sorts of constructs were removed or added?
This can reveal patterns of code that were thread unsafe in the pre-multicore
code that developers (especially those not participating in the multicore port)
should be made aware of in the future.  It can also reveal repeated patterns that
were added, indicating potential refactorings that may be desirable to apply
in order to reduce the proliferation of code clones within the project.


Language designers may want to know whether specific syntactic constructs would
make the language more productive for users. Taking an example from Java, we
might consider the addition of the for-each loop construct. This feature
could be partially justified by doing an analysis of existing source code to
determine that most for-loops iterate over an entire collection. To strengthen
this argument, it would be insightful to know what is the impact of maintaining
the code without for-each. For example, if refactoring the code commonly
leads to editing the bounds to match the collection used, then the argument in
favor of adding for-each is strengthened, as now it helps to prevent a
class of bugs where programmers forget to update the bounds.

Software developers joining a new project or team are expected to learn the
source code that they will be working with. We would like to provide these
programmers with tools that aid them in this task by allowing them to see what
types of changes other team members have made in the past. Software developers
may also want to compare the changes that happen in response to related bugs,
hoping to find opportunities to improve software quality, either by searching
for buggy patterns in the source code or making a tool to detect the pattern in
new code.  

\subsection{Related work}

The use of version control repositories as a source of data to study changes
to code over time is not new, but our approach to the problem is novel.
Neamtiu~\cite{neamtiu05understand} uses a similar approach of analyzing the
abstract syntax tree of code in successive program versions, but focuses on
detecting change occurrences only instead of going a step further and
attempting to identify any common patterns of change that can be found.  
Other groups have focused on identifying patterns based on common
refactorings that can be identified in the code~\cite{weissgerber06identify},
and seek to infer simple abstract rules that encapsulate the changes
that they detect~\cite{kim07automatic}.  For example, one such rule could
indicate that for all calls that match a certain pattern, an additional
argument should be added to their argument list.

This goal of generating abstract rules is similar to our goal of inferring
generic patterns in terms via
antiunification~\cite{reynolds70antiunification,plotkin70antiunification}.
What differs with our approach is that we presuppose no knowledge of the
underlying language beyond the structure provided by the language parser and
its mapping to an annotated term (or, aterm)~\cite{brand00aterm} format.  As
such, it is challenging to build rules that give an interpretation to the
program abstract syntax, such as ``append an argument to the function call'',
since we do not provide a mapping from the concept of ``function call'' to a
pattern of AST nodes.  By instead emitting templates in terms of the language
AST in aterm form, we are able to keep the tool as language-neutral as
possible.

\section{Methodology}
\label{sec:method}

\begin{figure}
\begin{center}
\includegraphics[height=0.44\textheight]{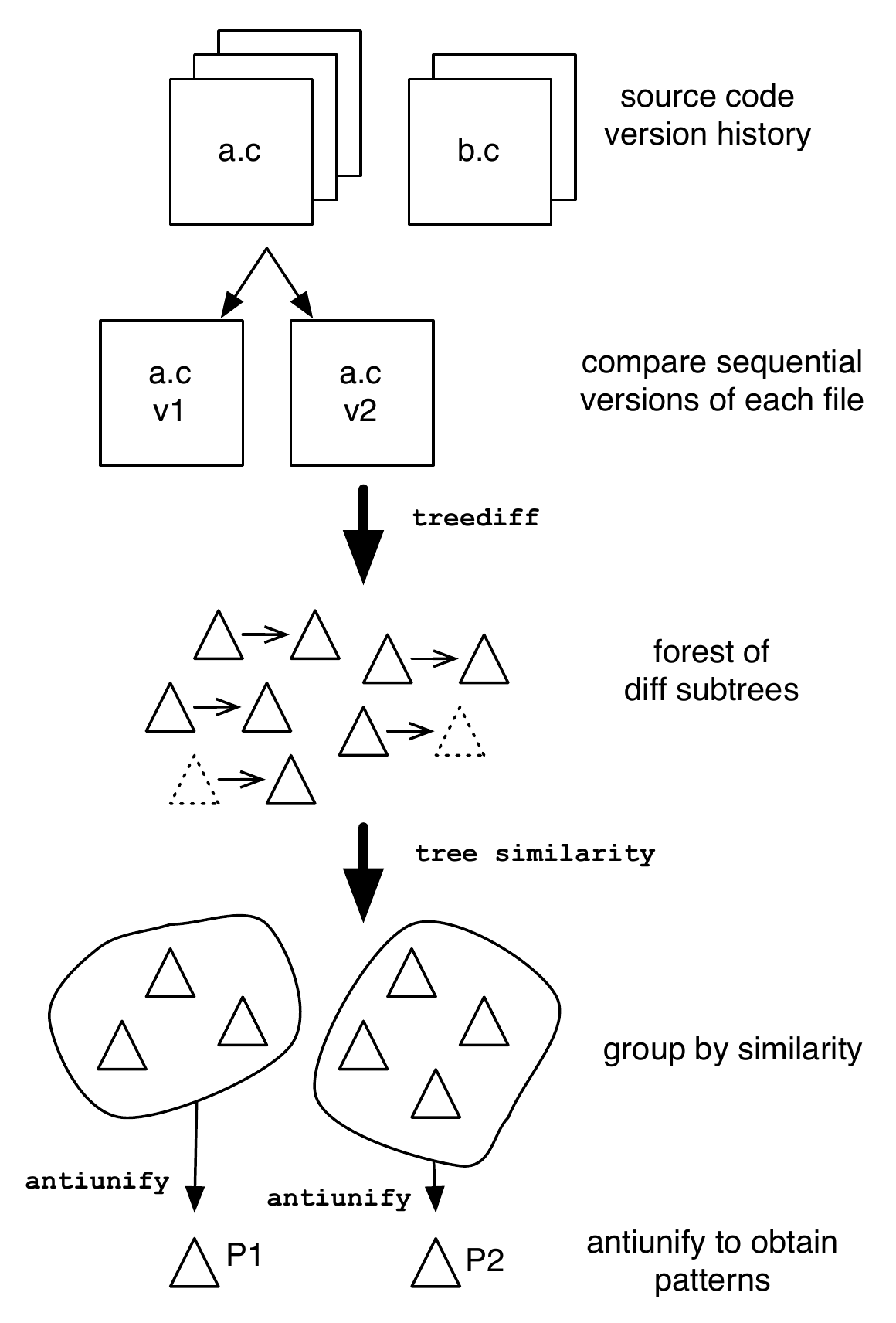}
\caption{The components of our prototype indicating how VCS data is
broken down into groupings of related changes for pattern generation.}
\label{fig:workflow}
\end{center}
\end{figure}

We propose the tool workflow illustrated in Figure~\ref{fig:workflow} for
studying software evolution via VCS data. First, each version of all source
files in the project are reconstituted from the differences stored within the
VCS such that each version of a file can be parsed by an appropriate language
front end.  Each front-end is configured to map the parsed code to an aterm
that represents a standardized serialization of the AST\@.  Mapping languages
to a common aterm format allows the downstream portions of our workflow to be
language-agnostic to a large degree, with minimal language-specific
parameterization.

Once we have code in an aterm format, we can then apply a structural
differencing algorithm between adjacent versions of each source file (e.g.,
version $n$ of file $f$ is compared to version $n+1$ of file $f$).  The result
of this is a forest of trees that represent the portions of the AST of file
$f$ that changed between versions at the structural level.  These changes can
either be code insertions, deletions, or mutations.  Our differencing is based
on the work of Yang~\cite{yang91diff} whose algorithm was designed for
computing differences between source code versions.  Yang's goal was to
improve the visual presentation of differences in textual diff tools, and our
use of their algorithm to provide input to further tree analysis algorithms is
novel.

After reducing the sequence of differences stored in the VCS, we have a large
forest of trees each representing a change that occurred over the evolution of
the software.  At this point, we seek to relate each of these trees via a
tree similarity metric.  This is achieved by using Yang's algorithm a second
time, but in this case we ignore the sequence of edit operations that it
produces and simply consume the quantitative similarity metric that it
produces as a rough estimate of how closely related two trees are.  A
threshold parameter is defined in which two trees with a similarity above the
threshold are considered to be part of the same group of difference tress.

Finally, once the set of differences are grouped into groups of trees that are
similar up to the threshold, we perform antiunification on the entire group to
distill all members to a representative code pattern for the group.
Antiunification of a set of terms yields the least general generalization of
those terms, which is how we define our notion of a code pattern.  The
antiunification algorithm as described by Bulychev~\cite{bulychev08dupe} as
part of the {\it clonedigger}
project\footnote{\url{http://clonedigger.sourceforge.net}} was used, which
itself is an implementation of the classical antiunification algorithm
described by both Reynolds~\cite{reynolds70antiunification} and
Plotkin~\cite{plotkin70antiunification}.

In the following sections, we describe the steps above in greater detail.

\subsection{Parsing and aterm generation}

One of the most challenging aspects of performing this kind of study on
arbitrary software packages is the availability of robust language parsing
tools.  In the absence of a common intermediate representation or abstract
syntax representation for popular languages, we adopted a standardized
serialization format in the form of annotated terms.  Generation of aterms was
achieved via language-specific parsers.  In this work, we used the
{\tt language-java} parser available as an open source library accessible via the
Haskell programming language.

The structure of aterms is given by this simple syntax:
\setlength{\grammarindent}{8em}
\begin{grammar}
<aterm> ::= `AAppl' <string> <aterm-list>
\alt `AList' <aterm-list>
\alt `AInt' <int>

<aterm-list> ::= <aterm> <aterm-list>
\alt $\epsilon$
\end{grammar}

This structure is sufficient for us to encode typical abstract syntax trees if
we allow ourselves to use the string label of the {\tt AAppl} portion of the
aterm. This is most easily illustrated with an example.  Suppose that we have
the Java AST for the statement {\tt i++;}.  In a textual form, this portion of
the AST would be represented by:

\begin{verbatim}
ExpStmt
  (PostIncrement
    (ExpName
      (Name [Ident "i"])))
\end{verbatim}

The translation to aterm would give us:

\begin{verbatim}
AAppl "ExpStmt"
  [AAppl "PostIncrement"
    [AAppl "ExpName"
      [AAppl "Name"
        [AAppl "Ident" [Appl "\"i\"" []]]]]]
\end{verbatim}

Notice that for strings, such as identifier names, we place double quotes
around the string inside the label portion of the aterm. Implementations of
aterms often provide a representation that allows for nodes to be shared
within the tree. While this is a useful optimization for saving space, we
chose to use the simpler unshared representation in our prototype due to the
clearer expression of the tree analysis algorithms over the unshared form of
the structure.

\subsection{Structural differencing}

One of the classical algorithms studied in computer science is that of string
similarity and the concept of string edit distance as a measure of the minimal
number of operations necessary to mutate one string or sequence into another.
A more complex problem is to define a similar sequence of operations to change
a non-linear structure like a tree from one into another.  This problem of
computing a structural edit distance has been studied since the 1970s and has
yielded tree differencing algorithms analogous to string differencing
algorithms commonly used in text analysis.  Many modern efforts in this area
are based on the initial work of Selkow~\cite{selkow77tree} and
Tai~\cite{tai79tree}.  Interest in such analysis of tree-structured data
increased with the proliferation of structured document formats used on the
Internet such as XML, HTML and SGML (a noteworthy example from this body
of work is found in Chawathe~\cite{chawathe96change}).

Our work is based on Yang's source differencing technique~\cite{yang91diff}.
In this algorithm two trees to be compared are mapped to two trees of edit
operations in which nodes from the original trees are annotated with edit
operations ({\it keep} or {\it delete}).  These can be applied to turn each tree into the
other.  On their own the edit trees are not sufficient to identify the paired
subtrees that represent regions where change occurred.  This requires an
additional step of processing the edit trees to form a single tree in which the
edit trees have been woven together.

\subsection{Identifying structural changes via edit tree weaving} 
\label{sec:weaving}

Ideally, we would like to obtain from the tree differencing algorithm what can
be thought of as the two trees overlaid on each other such that the common
structure from the root towards the children is clear, and points where subtrees
differ are explicitly identified.  The details on how this algorithm was 
implemented are not critical to this paper --- instead, we will focus on what the
woven trees contain.  In the discussion that follows, we adopt the 
convention that the arguments to the binary tree differencing function are
referred to as the \emph{left} and \emph{right} trees.  

Changes that occur between the trees are represented by three change types. If
the difference between two trees is the insertion of a subtree in the right
tree, then the woven tree will contain a \emph{left-hole}.  Similarly,
deletion of a subtree from the left such that it is not present in the right
tree will result in a \emph{right-hole}.  If a subtree was determined to be
changed, then the woven tree will contain a \emph{mismatch} point that refers
to the both the right and left subtrees that differ.  All other points in the
tree that match are joined with a \emph{match} point that contains the
corresponding common node to both trees.  




Given two edit trees that have been woven together into a tree with explicit
holes and mismatches, we can extract the subtrees that correspond to the three
types of changes above.  Match points also play an important role in
extracting changes by retaining the common context that was present in both
trees where the change occurred.  If we extract only the subtree rooted at the
point where the change occurred, the rest of the analysis will be missing the
context where the change took place. This information is necessary when
constructing understandable patterns.  

For example, while it may be true that a code fragment such as {\tt i++} is
where the change occurred, it is most useful to know whether or not that
fragment occurred within an expression, a for-loop, or as a standalone
statement. As such, we have chosen for the work presented here to extract the
subtree along with the closest enclosing statement. For example, if the
subtree was the expression {\tt i++;} within the statement {\tt if(i < 100)
i++;} we would extract the if-statement with the expression.

This is achieved by including the subtree rooted at the nearest ancestor
(which must be a matching point in the woven edit trees) to a change
representing an appropriate abstract syntax element.   In the future, we would
like to explore other ways of extracting context, such as looking at the
closest enclosing expression, function (when it exists), or class.  This
information should also be parameterizable to support differences in important
AST nodes that varies between languages.

\subsection{Tree similarity metric and grouping}

Given two trees $t_1$ and $t_2$, we would like to define a similarity metric
such that $d(t_1, t_2) \in [0,1]$, where a similarity of $1$ means that the
trees are identical, and $0$ represents maximal dissimilarity.  In Yang's
algorithm, a similarity score is provided for comparing $t_a$ and $t_b$. This
metric is order dependent, forcing the maximal score to be the size of the
left tree ($t_a$), even if $t_b$ is larger.  If the trees are identical, the
score will be exactly $|t_a|$, the number of nodes in $t_a$.  If they differ,
it will be strictly less than $|t_a|$.  As such, it would be possible to define
our distance function to be $\frac{d(t_a, t_b)}{|t_a|}$, but this operator is
not symmetric, since it is easy to find instances such that $\frac{d(t_b,
t_a)}{|t_b|} \neq \frac{d(t_a, t_b)}{|t_a|}$ when the trees are very different.
Instead, we define $\Delta(t_a, t_b)$ to be the function
$$\Delta(t_a, t_b) := \frac{min(d(t_a, t_b),d(t_b, t_a))}{max(|t_a|,|t_b|)}$$
where the $min$ and $max$ functions force the calculation to be symmetric.

\begin{figure}
\begin{center}
\includegraphics[width=0.45\textwidth]{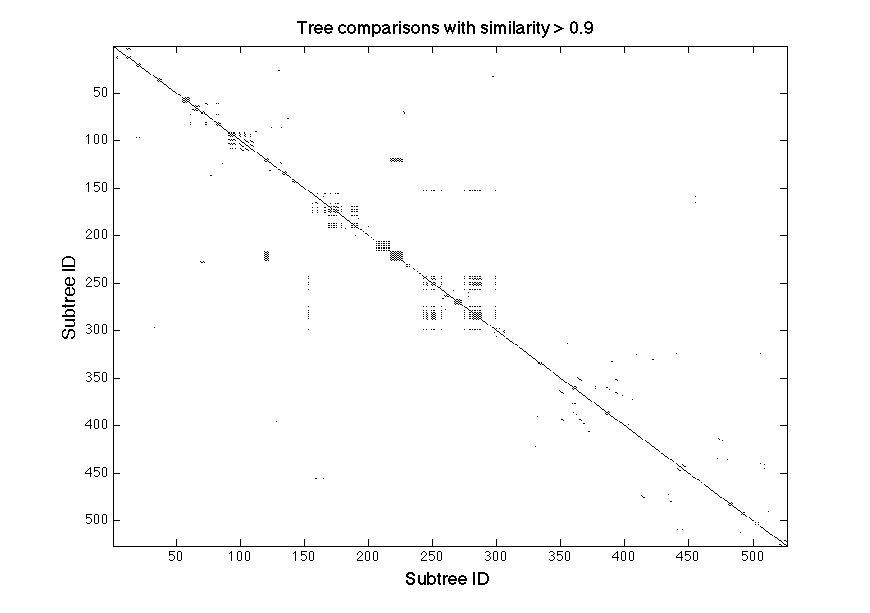}
\caption{Boolean matrix $D$ for over 500 changes from the ANTLR repository indicating all pairs of
changes for $\tau = 0.9$.}
\label{fig:boolmat}
\end{center}
\end{figure}

Once we have the set of changes that were detected from the VCS history, we
can generate a forest of trees $t_1, \cdots, t_n$ obtained from the holes and
mismatch points in the woven edit trees.  We then compute the $n^2$ distances
between all pairs to generate a distance matrix $D$ where $D_{ij} =
\Delta(t_i, t_j)$.  Given a threshold value $\tau$, we can produce a boolean
matrix $D'$ where $D'_{ij} = \Delta(t_i, t_j) > \tau$.  An example matrix is
shown in Figure~\ref{fig:boolmat} for changes observed in the VCS for ANTLR
where $\tau = 0.9$.  Note that for large numbers of changes, a sparse
representation of the boolean matrix can be computed for a given $\tau$
without requiring the full dense distance matrix to be created.  The sparsity
of the matrix is dependent both on the types of changes present and the value
of $\tau$ chosen.  

In our implementation, we create multiple distance matrices such that each
represents only related changes of a certain type from the woven tree (left
and right holes, and mismatches).  The matrix as defined above is simply the
element-wise boolean or of these three matrices.  Capturing this information is
important as it allows us to further refine our view of the code evolution to
distinguish code changes from the insertion or removal of code that occurs
over time. For example, when code is being developed and grown, we expect to
see a number of code insertions. Similarly, when a mass refactoring occurs to
simplify code, we would expect to see a set of code deletions.  When a more
subtle refinement occurs, such as transposition of code arguments or the
addition of a conditional to refine control flow, we would expect to see
mismatches where the tree changes.


\subsection{Antiunification and template generation}

Once we have groups of related code snippets in the form of related subtrees,
we can seek patterns that relate changes.  For example, say we have a function
call {\tt foo()} where each invocation of the function uses the same
parameters (e.g, {\tt foo(x,y)}, where {\tt x} and {\tt y} are always the
same). If we add a new parameter at the end of each call where the variable
passed in differs each time (e.g., {\tt foo(x,y,a)} and {\tt foo(x,y,b)}), we
would like to abstract out this change as {\tt foo(x,y,\metavar)}, where each
instance of the change replaces \metavar~with whatever concrete variable is
used at that point. The antiunification algorithm is built for this purpose --
given two trees, it seeks the least-general generalization of the pair and
produces a triplet representing the generalized tree with a metavariable
introduced where the two differ, as well as a substitution set that allows the
metavariable to be replaced with the appropriate concrete subtree necessary to
reconstitute the two trees that were antiunified.  Multiple distinct
metavariables (\metavar$_1$, $\cdots$, \metavar$_n$) are used when multiple
independent \metavar~points are necessary to represent a generalized set of
trees.

\section{Experimental results}

We tested the methodology outlined in Section~\ref{sec:method} on the
publicly available git repositories for two popular open source
projects, the ANTLR parser generator and the Clojure language implementation.
Both are implemented in Java, and one (ANTLR) is composed of a mixture of
hand-written and automatically generated code.  

\subsection{Threshold sensitivity}
\label{sec:threshold}

The first experiment that we performed was to investigate the effect of
similarity threshold to the number of groups identified, as well as the degree
of generality present in the tree that results from all members of each group
being antiunified together. Our prediction was that at the lowest threshold
($\tau = 0.0$), when all trees are considered to be similar, their antiunification will
yield the most general pattern.  This is what was observed, in which the
antiunification result is a tree composed of a single metavariable node.
Similarly, at the highest threshold ($\tau = 1.0$), the only groupings that will be
present will be single tree sets, or sets containing identical trees for
instances of identical changes that occurred in different places.  This is
precisely what we observed, with the antiunified trees containing no 
meta-variables since antiunification of a set of identical elements is the 
element itself.

\subsection{Group counts}
\label{sec:groups}
  
\begin{figure}
\begin{center}
\includegraphics[width=0.44\textwidth]{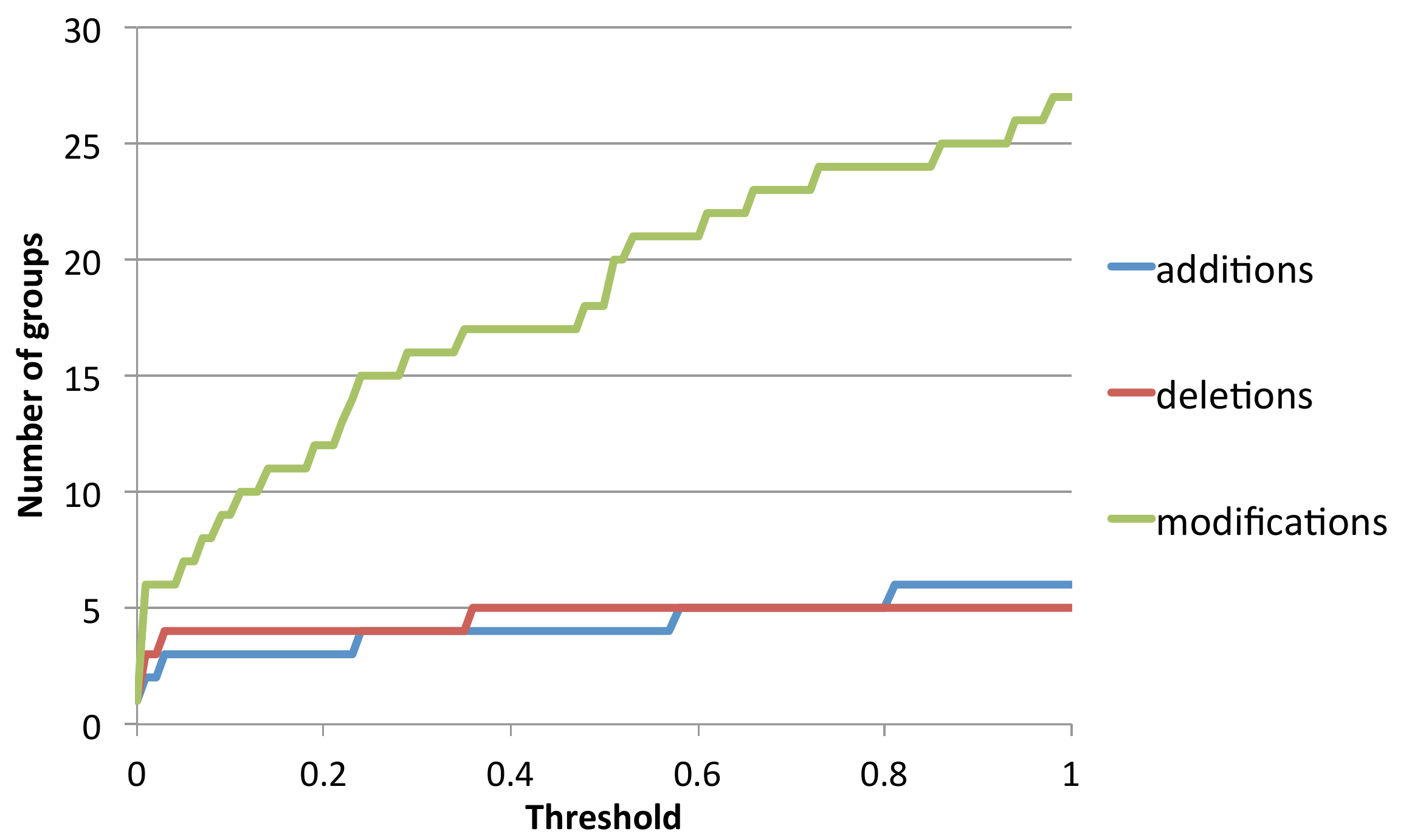}
\caption{Number of additions, deletions, and modifications by threshold for the Clojure source}
\label{fig:clojure-number-of-modifications}
\end{center}
\end{figure}

\begin{figure}
\begin{center}
\includegraphics[width=0.44\textwidth]{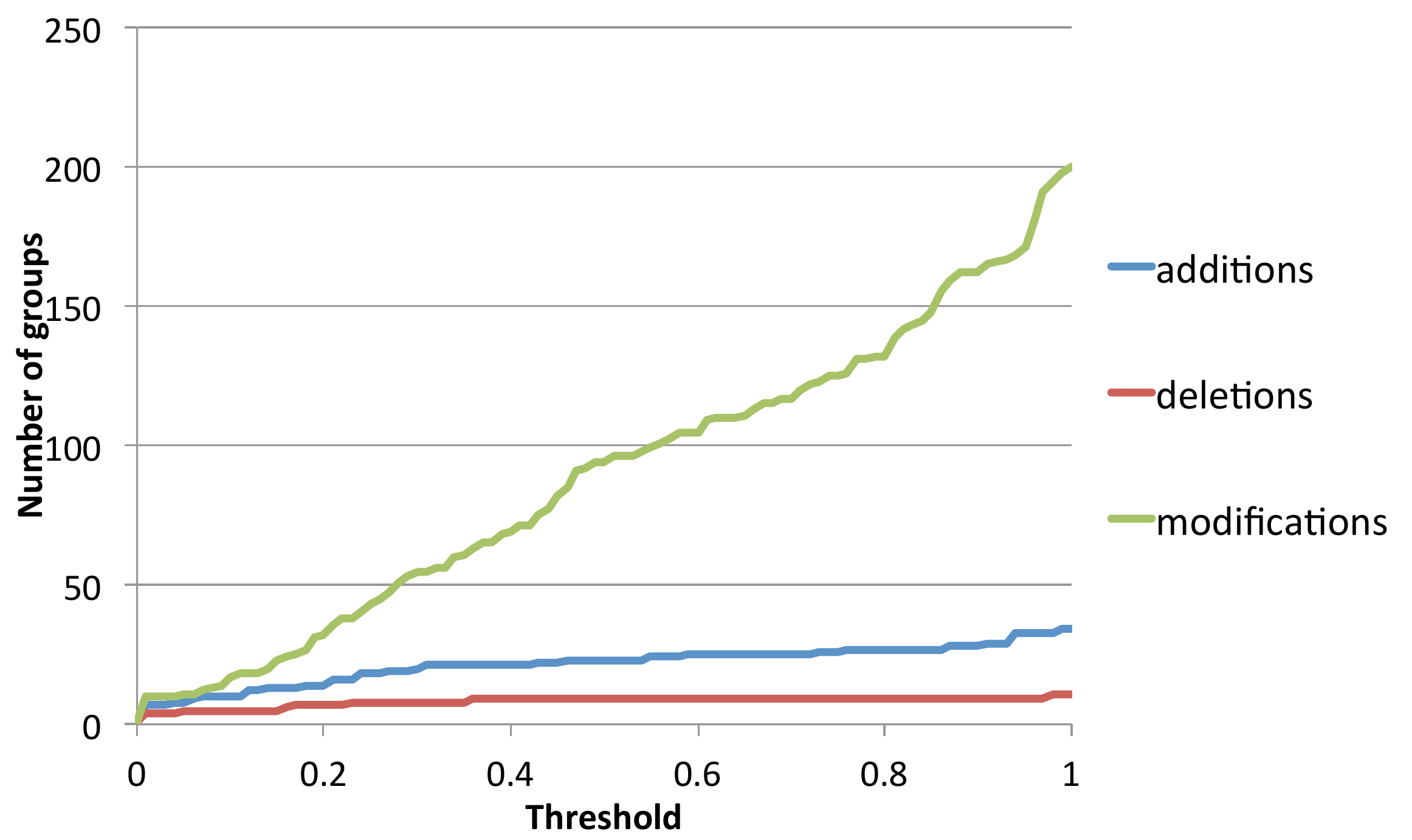}
\caption{Number of additions, deletions, and modifications by threshold for the ANTLR source}
\label{fig:antlr-number-of-modifications}
\end{center}
\end{figure}

We show the number of groups (broken down by type: addition, deletion, or
modification) as a function of threshold of similarity ($\tau$).
Figure~\ref{fig:clojure-number-of-modifications} shows the number of groups
for the Clojure history and Figure~\ref{fig:antlr-number-of-modifications}
shows the number of groups for the ANTLR history. In both cases, we only
consider a small portion of the full history of the VCS.

At the maximum $\tau = 1.0$, the total number of changes is less than the
number of trees we started with, because some changes end up being identical.
As we can see, as $\tau$ increases, we see more groupings of changes due
to changes that were considered similar under a lower threshold being
considered dissimilar under the more restrictive threshold.  Increases in
the group count represent large groups splitting into one or more smaller groups.

As an example, at $\tau=0.15$, a single pattern for for-loops is identified:


\begin{java}
for ($\metavar$ = $\metavar$ ; $\metavar$ < $\metavar$ ; $\metavar$) {
    $\metavar$
}
\end{java}

As the threshold is increased to $\tau=0.25$, in addition to generic for-loops, a
cohort of changes are identified to a more specific instance of the for-loop where
the loop counter is initialized to zero:

\begin{java}
for ($\metavar$ = 0 ; $\metavar$ < $\metavar$ ; $\metavar$) {
    $\metavar$
}
\end{java}

Increasing to $\tau=0.35$, the pattern for the conditional becomes more specific
and we see what appears to be a template for using the field of an object
(e.g., {\tt args.length}) as the loop termination criterion:

\begin{java}
for ($\metavar$ = 0 ; $\metavar$ < $\metavar$.$\metavar$ ; $\metavar$) {
    $\metavar$
}
\end{java}

Similar templates emerge for code patterns such as method invocations, printing
the concatenation of two strings, and other common activities.  


\subsection{Pattern identification}
\label{sec:clojure}


Using a portion of the Clojure history, we varied $\tau$ from 0 to 1 with an
increment size of 0.01 as shown in Figure~\ref{fig:clojure-number-of-modifications}.  
Looking at just the number of deletions, we examined the
point where the number of deletions goes from 4 to 5 as the threshold changes
from 0.35 to 0.36.

The following code, presented in standard style of unified diff, shows a loop
and the lines that were removed. This example comes from a file named {\tt
PersistentArrayMap.java}:


\begin{java}
 public Object kvreduce(IFn f, Object init){
     for(int i=0;i < array.length;i+=2){
         init = f.invoke(init, array[i], array[i+1]);
-           if(RT.isReduced(init))
-                   return ((IDeref)init).deref();
         }
     return init;
 }
\end{java}

Given the low threshold, this deletion was considered to be similar to the
example from {\tt PersistentHashMap.java} below.
Note that whitespace in most languages is syntactically neutral and curly braces
are optional for single statement conditional or loop bodies.  As a result,
the parser used in this work gives the same AST for \verb|if (exp) { stmt; }|
and \verb|if (exp) stmt;|.   Such changes are intermingled with syntactically meaningful
changes in the unified diff format.  To clarify the specific difference that our tool 
considers to actually be different, we have added a ``>'' prefix to the appropriate lines
of the unified diff.

\begin{java}
 public Object kvreduce(IFn f, Object init){
-    for(INode node : array){
-        if(node != null){
+    for(INode node : array)
+        {
+        if(node != null)
             init = node.kvreduce(f,init);
>-                if(RT.isReduced(init))
>-                        return ((IDeref)init).deref();
-               }
-           }
+        }
     return init;
 }
\end{java}

In both cases, our tool identified for-loops where the same lines are removed.
In fact, the code for both of these is very similar perhaps owing to Java's
HashMap and ArrayMap classes being very similar in terms of interface.
Furthermore, it did this at the statement level, eg., we did not need to
consider the similarities of the file names or the method names.  The jump in
group count as $\tau$ increased corresponds to the differences in the for-loops 
that contain the change falling below the necessary threshold of
similarity for grouping.




\section{Conclusions and future work}

We have shown that patterns of change over the lifetime of a project can be
obtained through analysis of its version control history.  The use of tree
differencing and tree similarity measures, as well as the antiunification
algorithm for computing generalized patterns, allows this large volume of
difference data to be distilled into a compact form in which changes can be
studied at the level of the base language syntax.  Analysis of the size and
count of groups of similar changes as a function of a similarity threshold
provides a disciplined way to identify generalizations of changes identified by
the tool.

Our work has been performed using a generic, language neutral term
representation allowing the same techniques to be applied to other languages
given appropriate parsing infrastructure and a mapping from language-specific
abstract syntax forms to the generic annotated term form.  Minimal
parameterization of the tool is necessary to then consume these terms, with
language-specific parameters largely focused on specific nodes within the term
that correspond to semantically useful subtree roots for providing context to
tree differences.

In Section~\ref{sec:threshold}, we showed that our approach can highlight the
evolution of code structurally. In fact, our example of the for-loop precisely
supports the hypothetical language designer argument laid out in
Section~\ref{sec:motivation}.

In Section~\ref{sec:clojure}, we were able to find related changes in different
files that happened as part of the same commit. Not only were we able to remove
noise compared to line-based diff, but we were also looking at the Clojure
source for the first time and able to see an important relationship between the
internals of the classes in those two files. As programmers who are completely
new to the Clojure source we were able to gain valuable insight.


Our experiment relied on a simple replay of the history of a software project.
There are other meaningful ways to generate the set of files to analyze. One
such example would be to correlate code changes to bug fixes and bug reports
and then push those changes through our workflow to find patterns.  As
mentioned in Section~\ref{sec:motivation} this may provide a support to quality
assurance practices.

In Section~\ref{sec:weaving} we explored one way to extract context. Many
different heuristics would be suitable here. Studying the trade-offs of
different heuristics would allow us to fine tune our approach depending on the
application and what we wanted to learn about the source code.

\subsection{Acknowledgments}

This work was supported in part by the US Department of
Energy Office of Science, Advanced Scientific Computing Research
contract no. DE-SC0004968.  Additional support was provided by Galois,
Inc.



%
\bibliographystyle{abbrv}
\bibliography{sigproc}  
\end{document}